# PERCUTANEOUS RENAL PUNCTURE: REQUIREMENTS AND PRELIMINARY RESULTS

A. LEROY[1], P. MOZER[1,2], Y. PAYAN[1], F. RICHARD[2], E. CHARTIER-KASTLER[2], J. TROCCAZ[1]

[1]Laboratoire TIMC - Faculté de Médecine - Domaine de la Merci 38706 La Tronche cedex
[2]Service d'urologie et de transplantation rénale. CHU Pitié-Salpêtrière. AP-HP 75013 Paris
Author for correspondence: A. Leroy, Antoine.Leroy@imag.fr

## INTRODUCTION

Percutaneous access to kidney is a challenging technique that meets with the difficulty to reach rapidly and accurately an intra-renal target. Today, puncture guidance is performed under fluoroscopic or echographic imaging, each of which presents drawbacks: fluoroscopy provides limited 2D information on localization, whereas echography mostly gives fuzzy images of the target and the puncture trajectory [6].

This paper introduces the principles of computer assisted percutaneous renal puncture (PRP), that would provide the surgeon with an accurate pre-operative 3D planning on CT images and, after a rigid registration with space-localized echographic data, would help him to perform the puncture through an intuitive 2D/3D interface.

The whole development stage relied on both CT and US images of a healthy subject. We carried out millimetric registrations on real data, then guidance experiments on a kidney phantom showed encouraging results of 4.7mm between planned and reached targets.

## PRE-OPERATIVE PLANNING

### 1. CT Images Acquisition

Modern CT scanners can provide high-quality images. We acquired two exams of a healthy volunteer, the voxel size being submillimetric (0.6x0.6x0.6). Such an exam is systematically performed on the patient before intervention, thus does not induce more irradiation than necessary.

The first CT-exam was performed early after the injection of a contrast product, to highlight the renal cortex, whereas the second exam, 5min later, gives accurate information on the Pyelo-Calicial Cavities (PCC) (often the target to reach). See fig. 1.

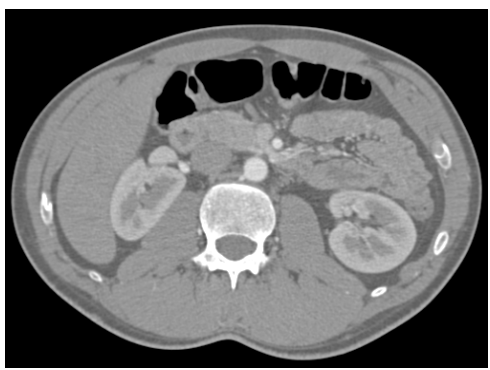

**fig. 1: Early CT acquisition. The renal cortex is enhanced**

*2. CT Images Segmentation*

The CT images were segmented using derivatives methods (Nabla's 3D watershed, Generic Vision), which we found far more accurate anatomically than the morphological operators provided by Analyze (BIR, Mayo Clinic).

Relying on the kidney contours in both CT exams, the segmented PCC in exam2 were then registered to exam1 with an accuracy of 1mm (using Analyze volume registration). Therefore, both external and internal kidney structures were available in a unique CT coordinate system, for a more accurate planning (fig. 2).

The last stage in our pre-operative segmentation was to export the segmented structures as 3D meshes (typically the skin, rachis, kidneys and the registered PCCs). The reference structure used for intra-operative multimodal registration is the kidney surface.

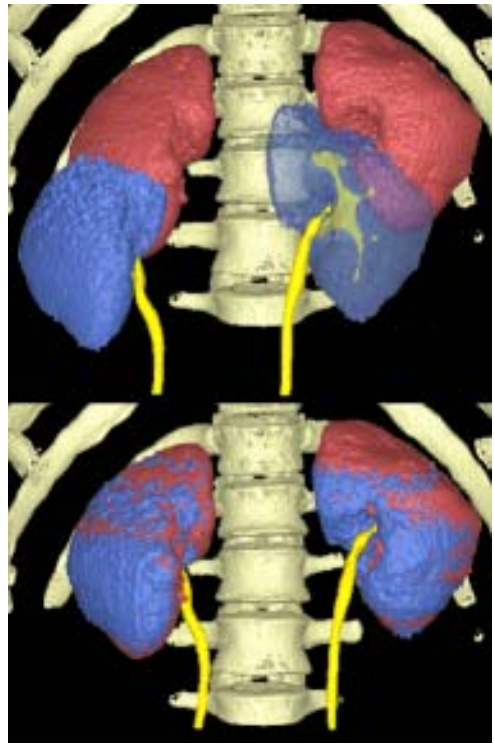

**fig. 2: PCCs in exam2 are registered to exam1**

*3. Pre-Operative Planning*

The planning phase allows the selection of 2 points, a target and a source, which define the needle trajectory. The selection is performed on 2D ortho and oblique slices, and on the 3D scene, for improved usability.

## INTRA-OPERATIVE REGISTRATION

*1. Echographic Images Acquisition*

The acquisition was carried out on a Hitachi-EUB405, the probe (3.5MHz for abdominal structures) being localized in space thanks to a "rigid body" mounted on, and to the Polaris system (NDI) [1,7]. The estimated time elapsed between the recordings of the rigid transformation and of the image is 70ms, which induces an error of 0.7mm at a 10mm/s motion.

Furthermore, we noticed that the anterior access for echographic acquisition was more appropriate to get good-quality images and detailed segmentations, although most PRP are done through posterior access.

The acquired kidney remains in place, since the subject holds his breath, as long as the acquisition does not last more than 30-40 seconds (under global anaesthesia, the control of the breathing should be easier, thanks to the breathing device and to the lack of stress).

We generally acquire 200 images at 3 images/s, in both transversal and longitudinal orientations. All images are not segmentable, in fact.

*2. Echographic Images Segmentation*

In that feasibility study, the kidney cortex was segmented manually on the echographic images. The intrinsic and extrinsic calibrations of the echographic probe [3] allowed to replace the 2D-segmented points into 3D space (this is called "2.5D echography" [1,7]). Fig. 3 shows a segmented longitudinal kidney.

We noticed that a dense, homogenous cloud of points (fig. 3) was suitable for the registration phase. However, The user may also focus on the structures close to the target, and also on some high-curvature regions (fig. 4), that will avoid local minima during registration.

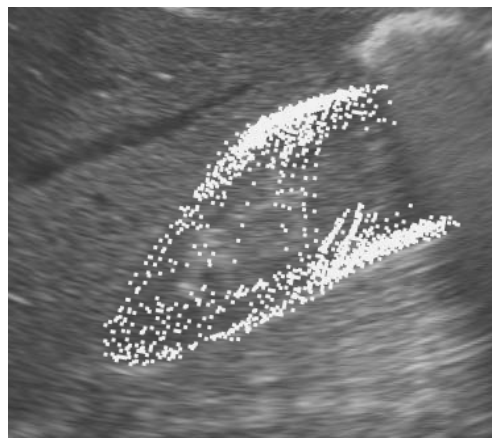

**fig. 3: Segmented right kidney**

*3. 3D/3D Rigid Registration*

Unlike the liver [5,7], the kidney cortex is fairly hard, so we chose to perform a rigid registration of the pre-operative planning into the intra-operative coordinate system. An ICP algorithm based on octree-splines and Levenberg-Marquardt minimization [4] matches the echographic cloud of points of the cortex onto the CT mesh (fig. 4).

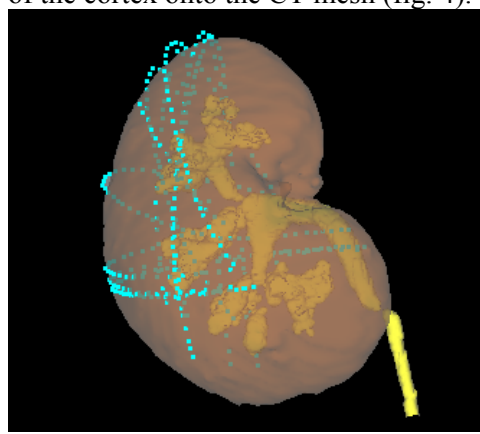

**fig. 4: Pre- and Intra- operative right kidneys matched**

# VALIDATION: ACCURACY AND PUNCTURE TESTS

## 1. Registration: Repeatability Tests

Tab. 1 shows the results obtained for 6 initial positions. A transform is represented as one translation vector and three rotation angles. The deviations between the final position and the 6 initial attitudes go up to 30 mm in translation and 20° in rotation. Beyond those values, local minima are quasi-systematically found. One can see that the results are fairly good.

|     | Test 1 | Test 2 | Test 3 | Test 4 | Test 5 | Test 6 | Mean  | σ   | ‖σ‖  |
|-----|--------|--------|--------|--------|--------|--------|-------|-----|------|
| Tx  | 270,2  | 268,8  | 272,7  | 270,5  | 272,9  | 276,1  | 271,9 | 2,6 | 1,0  |
| Ty  | 466,0  | 464,9  | 462,0  | 464,6  | 463,2  | 462,8  | 463,9 | 1,5 | 0,3  |
| Tz  | -332,3 | -335,1 | -332,9 | -333,2 | -332,1 | -327,6 | -332,2| 2,5 | 0,8  |
| ψ   | -85,2  | -86,3  | -83,0  | -85,6  | -83,4  | -82,1  | -84,3 | 1,7 | -2,0 |
| θ   | -44,5  | -45,0  | -41,8  | -44,3  | -42,2  | -41,0  | -43,1 | 1,7 | -3,9 |
| φ   | -179,7 | -179,8 | -178,7 | -179,9 | -179,2 | -179,8 | -179,5| 0,5 | -0,3 |

**tab. 1: Repeatability test results ( $\|\sigma\|=\frac{\sigma}{Mean}$ )**

## 2. Registration: Closed-Loop Tests

Let $CT_1$ and $CT_2$ be two CT meshes, where $CT_2$ is a transformed $CT_1$ (e.g. 50mm in translation and 10° on each rotation angle). Let $US$ be a US cloud of points of the same organ. $M_{12}$, $M_{1U}$ and $M_{2U}$ are the mono- or multi-modal transforms betweens the exams (fig. 5).

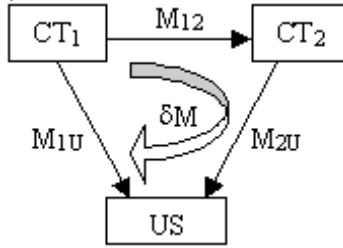

**fig. 5: Closed-Loop Test**

Our closed-loop test consists in evaluating $\delta M = M_{1U}^{-1} * M_{2U} * M_{12}$

The registration is perfect if $\delta M = Id$. Our results are:

$\|\delta M - Id\|_T = 5.9mm$ ; $\|\delta M - Id\|_R = 0.006$
$\|\delta M * CT_1 - CT_1\| = 1.2mm \pm 0.4$

## 3. Guidance: Puncture of a Phantom

This is the final accuracy test. Six trajectories were planned on the abdominal phantom, 3 for each kidney. Fig. 6 shows one of those, and the puncture needle.

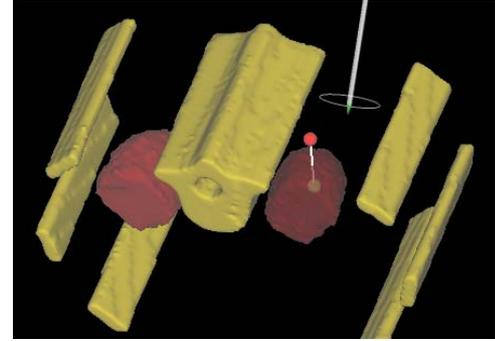

**fig. 6: Planned trajectory and needle on right kidney**

We used urological needles, which are very soft (0.9mm wide, 200mm long). Despite a painful pre-operative segmentation (the phantom is made out highly heterogeneous material), registration and puncture succeeded.

We checked the position of the needles using echography: fig. 7 shows that the 2 first needles reached the PCC.

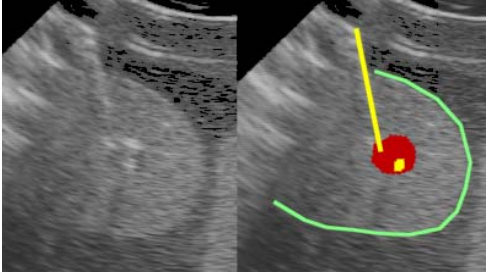

**fig. 7: The 2 first needles reached the PCC**

To get more accurate quantitative position assessments, the phantom had a post-operative CT exam, which showed that the needles were, on the right kidney side, 4.7mm away on average from their target (tab. 2). Log files also gave information on the accuracy of both the system - the most important for us at present - and the surgeon (respectively 2.1mm and 4.3mm).

On the left side, a local minimum in the registration is responsible for a needle-to-target error of 1cm, although the surgeon was very efficient in following the trajectory on the 2D/3D interface. Those kidneys were half-kidneys; we expect that, dealing with entire kidneys, the registration will be more reliable.

|            | P1  | P2  | P3  | Avg |
|------------|-----|-----|-----|-----|
| \|Pre-Post\| | 6,1 | 3,3 | 4,7 | 4,7 |
| \|Log-Post\| | 2,5 | 1,9 | 2   | 2,1 |
| \|Log-Pre\|  | 5,1 | 2,9 | 5   | 4,3 |

**tab. 2: Distances between pre-op target, post-op and logged needle positions, for 3 punctures on right kidney**

## DISCUSSION

### 1. Accuracy Issues

Many sources of error can be mentioned to explain our results. Apart from casual local minima during registration, we believe that a large part of the final error comes from the echography:

- The probe calibration, as described in [3], is not yet optimal; we obtained 2mm as maximum rms error.
- As said above, there is a small time gap between the probe localization and the image acquisition, resulting in a 1mm error at slow speed.
- The echographic segmentation, even manual, is not easy, as the interface between two structures is always difficult to locate with precision. We once experienced a 1mm translation during palpation due to a bad echographic segmentation.

But the major difficulty, to date, lies in the softness of the puncture needles. The system is reliable provided the calibrated tool remains rigid, thus inexperienced users might encounter finals errors over 10mm. Whereas pelvis puncture using hard needles has been proven efficient and accurate [1], puncturing through soft tissues remains a challenge.

### 2. Clinical Applicability

Our system still lacks for automation, especially for intra-operative segmentation.

Another burning issue, which was not yet mentioned in this paper, concerns breathing. We do not know at present whether the breathing device can place the kidney in the same position for

US acquisition and guidance [2], or if we should implement a real-time tracking.

### 3. Puncture of a Cadaver

The puncture of the phantom was a first step in evaluating our CAS system. Two obstacles made it uneasy: one the one hand, segmenting the CT images was more difficult than with human data because of the heterogeneity of the material, and on the other hand, the phantom was incomplete, as it contained only the superior part of the kidneys (fig. 6), making the registration sometimes hazardous.

So we planned a first experiment on real tissues for September 2002. We aim at puncturing both kidneys of a cadaver, in spite of the low echogenicity of the dead tissues. Surgery will certainly be necessary, even for anterior access echography.

# CONCLUSION

In this paper, the bases of a computer-assisted system for percutaneous kidney puncture were presented. The aim was to evaluate the feasibility and the accuracy errors at each step of the process. In our study, pre-operative CT data were registered with intra-operative, manually segmented US data, using a 3D/3D rigid matching. Tests on registration as well as guidance experiments were satisfactory. Further work will be undertaken to improve efficiency and accuracy in calibration and segmentation, and to take breathing into account.

We would like to thank Pr Passagia, Pr Tonetti, Dr Renard-Penna and Dr Pradel, as well as Le Laboratoire Pierre Fabre, l'Association Française d'Urologie and Generic Vision for their help and support.